%% file: main.tex
\documentclass[5p]{elsarticle}

\usepackage{graphicx}
\usepackage{amssymb}
\usepackage{subfigure}
\usepackage{numprint}
\usepackage{hyperref}
\usepackage[x11names]{xcolor}
\usepackage{graphbox}
\usepackage{bm}
\usepackage{framed}
\usepackage{caption}

\journal{Icarus}

\begin{document}

\begin{frontmatter}


\title{Tomographic inversion of gravity gradient field for a synthetic Itokawa model}



\author[tau]{Liisa-Ida Sorsa\corref{correspondingauthor}}\cortext[correspondingauthor]{Corresponding author}
\ead{liisa-ida.sorsa@tuni.fi}
\author[tau]{Mika Takala}
\author[mps]{Patrick Bambach}
\author[mps]{Jakob Deller}
\author[mps]{Esa Vilenius}
\author[mps]{Jessica Agarwal}
\author[gedex]{Kieran A. Carroll}
\author[rob]{\"Ozg\"ur Karatekin}
\author[tau]{Sampsa Pursiainen}

\address[tau]{Tampere University, Faculty of Information Technology and Communication Sciences, PO Box 1001, 33014 Tampere University, Finland}
\address[mps]{Max Planck Institute for Solar System Research, Justus-von-Liebig-Weg 3, 37077 G\"ottingen, Germany}
\address[gedex]{Gedex Systems Inc., 407 Matheson Blvd. East, Mississauga, Ontario, Canada  L4Z 2H2}
\address[rob]{Royal Observatory of Belgium (ROB), Avenue Circulaire 3, 1180 Brussels, Belgium}

\begin{abstract}
This article investigates reconstructing the internal mass density of a numerical asteroid model using the gradient of a simulated gravity field as synthetic measurement data.  Our goal is to advance the mathematical inversion methodology and find feasibility constraints for the resolution, noise and orbit selection for future space missions. We base our model on the shape of the asteroid Itokawa as well as on the recent observations and simulation studies which suggest that the internal density varies, increasing towards the center, and that the asteroid may have a detailed structure. We introduce   randomized multiresolution scan algorithm which might  provide a robust way to cancel out bias and artifact effects related to the measurement noise and numerical discretization. In this scheme, the inverse algorithm can reconstruct details of various sizes without fixing the exact resolution {\em a priori}, and the randomization minimizes the effect of discretization on the solution. We show that the adopted methodology provides an advantageous way to diminish the surface bias of the inverse solution. The results also suggest that a noise level below 80 Eotvos will be sufficient for the detection of internal voids and high density anomalies, if a sparse set of measurements can be obtained from a close-enough distance to the target. 
\end{abstract}


\end{frontmatter}


\section{Introduction}
\label{intro}
\input{introduction.tex}

\section{Materials and Methods}
\label{methods}
\input{methods.tex}


\section{Results}
\label{results}
\input{results.tex}

\section{Discussion}
\label{discussion}
\input{discussion.tex}

\section{Conclusion}
\label{conclusion}
\input{conclusion.tex}

\section*{Acknowledgments}

L.-I.S., M.T.\ and S.P.\ were supported by the Academy of Finland Centre of Excellence in Inverse Modelling and Imaging 2018-2025. 

\section*{References}




\bibliographystyle{model1-num-names}
\bibliography{refs.bib}







\end{document}

%% file: introduction.tex
Geophysical investigations into the subsurface of the Earth are nowadays based on combining information of multiple geophysical fields, leading to more reliable models of the subsurface structures \cite{erkan2011}. For example, subsurface cavities have been successfully detected by combining gravity field and ground penetrating radar measurements \cite{chromcak2016, mochales2008, panisova2013}. Furthermore, gravity gradiometry has been shown to detect local mass or density anomalies \cite{erkan2011, jekeli2011,  mckenna2016}. These investigations commonly have a low signal-to-noise ratio, hence the need to combine data from various measurement techniques. Such multi-modal approach is used to maximize the probability of anomaly detection and minimize that of a false alarm \cite{mckenna2016}.

The first attempt to investigate the deep interior structure of a small solar system body (SSSB) was the Comet Nucleus Sounding Experiment by Radio-wave Transmission (CONSERT), a part of European Space Agency's (ESA) Rosetta mission to the comet 67P/Churyumov-Gerasimenko, in which a radio signal was transmitted between the orbiter \emph{Rosetta} and the lander \emph{Philae} \cite{kofman1998, kofman2015, Kofman2007}. Other missions to asteroids have concentrated on the structure and composition of the surface of the target SSSBs. The Hayabusa mission \cite{hayabusa} by the Japanese Aerospace Exploration Agency (JAXA) explored the asteroid Itokawa extensively and measured the physical, chemical and geological properties of the body from orbit \cite{fujiwara, okada}. Furthermore, the Hayabusa mission returned a sample of the asteroid surface regolith for analysis on earth, confirming the classification of the asteroid to S-type \cite{nakamura, fujiwara}, as was originally reported by the earth-based visible and near-infrared spectroscopy observations \cite{binzel2001}. Analysis of the collected dust particles suggests that Itokawa is an asteroid made of reassembled pieces of the inner portions of a once larger asteroid \cite{nakamura, tsuchiyama}.  

The currently ongoing missions, Hayabusa2 to the asteroid 16173 Ruygu by JAXA \cite{tsuda} and OSIRIS-REx to the asteroid 101955 Bennu by NASA \cite{Lauretta2017} have been designed to measure the physical, chemical and geological properties of the target asteroids. 

Based on the experience obtained in terrestial applications \cite{erkan2011, chromcak2016, mochales2008}, the future missions to explore the interior structure of SSSBs would benefit from combining measurements from more than one geophysical field. Our recently published simulation studies of equipping CubeSats with a low-frequency stepped-frequency radar \cite{farfield, sorsa2019} suggest that tomographic reconstruction of the full electromagnetic wavefield can reveal internal structural anomalies in an SSSB. Therefore, augmenting such radar measurements with gravity field measurements by gradiometry presents an interesting opportunity to obtain complementary information on the structure of the target. Such a gravity gradiometer instrument has been suggested, for example, by the recent studies  \cite{carroll2018a, carroll2018b}. A combined gravity and radar measurement for interior investigation of an SSSB is, for example, a part of ESA's future mission plan HERA \cite{juventas}. The gravity field can be sensed via a  direct measurement by a gravity gradiometer \cite{carroll2018b} or indirectly by observing the Doppler shift of a radio signal transmitted by a spacecraft  \cite{konopliv2013jpl,andrews2013ancient}. The recent Moon Gravity Recovery and Internal Laboratory (GRAIL) mission used the latter method to measure the change in the distance between two co-orbiting spacecrafts as they flew above the lunar surface to calculate the gravitational potential from the spherical harmonics data set \cite{zuber2013}.

In this paper, we investigate inversion of simulated gravity measurements obtained for a synthetic asteroid model which is based on the shape of the asteroid Itokawa and augmented with interior density anomalies. Our  model relies on the recent observations and impact simulations which suggest that the internal density of asteroids varies, increasing towards the center, and that the interior may have a detailed structure in which void cavities and cracks have been formed in between rubble or larger boulders \cite{carry2012density,Jutzi2017,deller_hyper-velocity_2017}. Our goal is to advance the mathematical inversion methodology and find feasibility constraints for the resolution of the reconstruction, noise and orbit selection to guide the design of future space missions. 
We introduce and investigate a randomized multiresolution scan algorithm which might provide a robust way to cancel out bias and artifact effects related to the measurement noise and numerical discretization. In this scheme, the inverse algorithm can reconstruct details of various sizes without fixing the exact resolution {\em a priori}, and the randomization minimizes the effect of discretization on the solution. We show that the adopted methodology provides an advantageous way to diminish the surface bias of the inverse solution. The results also suggest that a noise level below 80 Eotvos will be sufficient for the detection of internal voids and high density anomalies, if a sparse set of measurements can be obtained from a close-enough distance to the target.

%% file: methods.tex
We use the gradient of the scalar  gravity field strength as input data  for our inversion procedure and study how the quality of the reconstruction depends on various parameters. We focus on the field strength for simplicity, assuming  that the gravitational torsion field, which is omitted here,  might involve more uncertainty, if the actual {\em in-situ}  measurement is done under a rotational motion. 

The two synthetic asteroid interior structure models have been created from the shape model of the asteroid Itokawa. The forward simulation of the field strength gradient is carried out in a uniformly regular tetrahedral finite element mesh \cite{braess2007finite}.  The mesh is generated with respect to a uniform point (vertex) lattice by subdividing each cube in the lattice into six tetrahedra. The synthetic  measurements are investigated for two orbit radii.  The sections \ref{sec:models}--\ref{sec:forwardmodel} describe the model and procedure used to create the simulated measurements.

In the inversion stage described in the section \ref{sec:inversionprocess}, we use a hierarchical Bayesian model which allows adjusting the hyperprior parameter for finding a well-localized reconstruction. The inverse estimate is found through a randomized multiresolution scanning technique in which the inversion mesh is decomposed to multiple, nested levels. Using this randomized scanning algorithm with Iterative Alternating Sequential (IAS) inversion algorithm, it is possible to average out discretization errors and hence the final inverse solution is less dependent on the discretization of the computation domain than it would otherwise be. 

\subsection{Asteroid models}\label{sec:models}

The asteroid models used in this work are based on the triangular stereolithography (STL) surface mesh \cite{itokawastlfile} of the asteroid Itokawa \cite{Saito2006}. The unstructured triangulated asteroid surface model was imported to Meshlab \cite{meshlab} and resampled with Poisson-disk sampling algorithm \cite{poissondisk} which produces a uniformly distributed set of points fulfilling a given minimum distance condition. This was done to obtain a regular surface mesh with a resolution comparable to that of the eventual volumetric mesh applied in the numerical simulations. 

The asteroid model consists of a surface layer and an interior compartment in which the deep interior anomalies are located. The surface of the  interior compartment  was created by further downsampling the shape model and rescaling it by the factor of 0.9. The same asteroid model structure was also used in  in radar simulations in \cite{sorsa2019}. 

\begin{figure*}[!ht]
\begin{center}
\begin{scriptsize}
   \centering
    \begin{minipage}{8cm}
   \centering    
    \includegraphics[width=4cm]{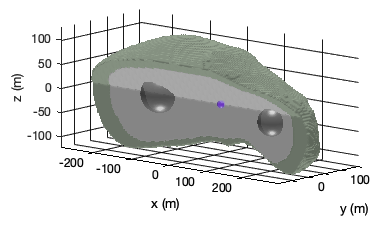}
    \includegraphics[width=3cm]{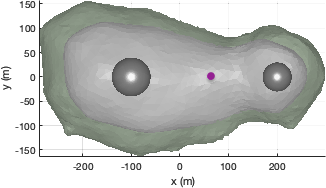}\\ Model (A) 
    \end{minipage}
     \begin{minipage}{8cm}
        \centering
    \includegraphics[width=4cm]{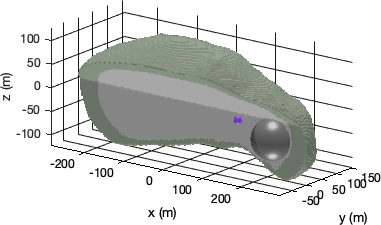} 
    \includegraphics[width=3cm]{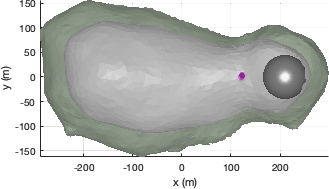}\\
     Model (B)
    \end{minipage}
    \end{scriptsize}
    \caption{3D cut views showing the structures of the exact models (A) and (B) with their centers of mass indicated by magenta. {\bf{Left:}} Model (A), the double void structure. {\bf{Right:}} Model (B), the high density spherical anomaly in the head of the asteroid.}
    \label{fig:exactmodel}
    \end{center}
\end{figure*}

The bulk density of Itokawa as measured during the Hayabusa mission is 1.9 $\pm 0.13$ g/cm$^3$ and its orthogonal axes are 535, 294, and 209 meters \cite{fujiwara}. A careful analysis of the rotational lightcurve observations and thermophysical analysis suggest that Itokawa is composed of two bodies with different bulk densities \cite{lowry2014}. A recent study on asteroid mass-concentration estimates obtained by asteroid impact simulations \cite{Jutzi2017} suggests that the bulk density of the deep interior part is higher than that of the surface. To simulate these features and test tomographic inversion, we created two interior models (A) and (B) depicted in the Figure \ref{fig:exactmodel}. 

In the model (A), the surface and the deep interior densities were adjusted to 1.8 g/cm$^3$ and 2.0 g/cm$^3$, respectively. This way, we could keep the bulk density close to the measured one while accounting for density variation between the surface and the deep interior structures. Two spherical void cavities of 40 and 30 meter radii and zero density were inserted into the asteroid body and the head, respectively. Based on the actual measurements of Itokawa by Hayabusa mission \cite{fujiwara} a rubble-pile asteroid may contain significant void space in the deep interior and these cavities model such anomalies. The locations were chosen so that one is in the body of the asteroid, in the deep interior part, and the other in the head, enclosed closer to the surface while being moderate in size. The center of mass of the model (A) is in $x = 64, y=0, z=0$ meters, assuming the centre of the coordinate system is in the geometrical center of the asteroid. 

In the model (B), the densities of the surface and the interior compartments were adjusted to 1.6 g/cm$^3$ and 1.8 g/cm$^3$, respectively. A spherical high-density  anomaly with a density of 8 g/cm$^3$, and 45 m radius was located in the head of the asteroid, covering approximately 55 \% of the total radius of the head and resulting in a total bulk density distribution similar to the findings in \cite{lowry2014}, and a total bulk density equal to that of Model (A), 1.9 g/cm$^3$. An internal  high-density anomaly within a rubble-pile asteroid could result for example from an impact event. The center of mass of the model (B) is in the coordinates $x = 122, y=0, z=-16$ meters.

\subsection{Measurement points}

\begin{figure*}[!ht]
\includegraphics[width=\textwidth]{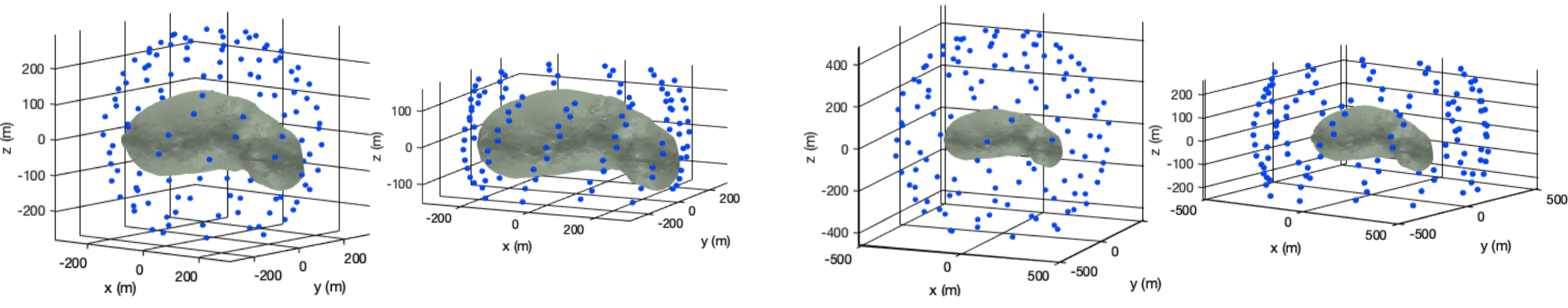}
\caption{{\bf{Left:}} Measurement radius 305 m from the geometrical center of the asteroid, resulting in a subset of points which are very close to the surface in the long axis, limiting angle (the angle between the ) 70$^\circ$ on the left, and 30$^\circ$ on the right. {\bf{Right:}} Measurement radius 500 m from the center of the asteroid. Limiting angles 70$^\circ$ on the left, and 30$^\circ$ on the right.}
\label{fig:measurementpoints}
\end{figure*}

The measurement point sets were modeled similar to our earlier work \cite{farfield, sorsa2019} in which two CubeSats orbit an asteroid at a defined radius measured from the geometrical center of the asteroid. Two circular orbits with radii of 305 m and 500 m were investigated. These orbits were selected to provide initial results on the quality of reconstructions that can be achieved when performing measurements and analysis with state-of-the-art instruments and tomographic inversion methods. While the lower orbit (305 m) is not realistic for performing satellite-based measurements, it was included to provide a reference for close-proximity measurements. The angular coverage between the measurement plane and the asteroid spin axis (here: spin around the z axis in the xy plane) was set to 70 or 30 degrees, resulting in the limited-angle spatial coverage of measurement points depicted in the Figure \ref{fig:measurementpoints}, with apertures around the z-axis. 

In practice, the gravity field measurements take time to carry out, resulting in measurement arcs rather than points. The effect of measurement arcs on the reconstructions was analyzed by introducing positional uncertainty in the model via rotating the measurement point set in comparison to the background model. Two rotation angles, 5$^\circ$ and 10$^\circ$, corresponding to realistic measurement times in the 500 m radius orbit (Table \ref{table:arcs}) were investigated. The orbit velocity was assumed to follow the equation $v=\sqrt{(GM/r)}$, in which $G$ is Newton's gravitational constant, $M$ the bulk mass of the asteroid and $r$ the orbit radius. The difference between the spacecraft orbit period and the spin rate of Itokawa (12.1 h, \cite{fujiwara}) resulted in approximately 5 \% positional uncertainty in the 500 meter orbit radius case. The difference between the spacecraft orbit velocity and the asteroid spin rate in the lower orbit is too large (99 \%) for a meaningful examination of reconstruction effects. 

\begin{table}[!ht]
    \centering
    \caption{The rotation angles and measurement durations resulting from realistic measurement arcs for the investigated orbits. The rotation angles are used in investigating the effect of measurement positional uncertainty in the inversion stage.}
    \begin{tabular}{lrr}
    \hline
        & 305 m orbit & 500 m orbit \\
        Rotation angle & Duration (s) &  Duration (s) \\ \hline
        5$^\circ$ & 610 s & 11814 s \\
        10$^\circ$ & 1220 s & 23628 s \\
        \hline
    \end{tabular}
    \label{table:arcs}
\end{table}

\subsection{Measurement noise}

We apply the recent noise estimate  \cite{carroll2018b} obtained for the VEGA space  gravimeter, when it is assumedly held by a 2.5 m boom attached to the mothership of the proposed HERA mission. Akin to \cite{carroll2018b},  the measurement noise distribution is assumed to be a Gaussian zero-mean random variable to account for several independent, identically distributed sources of noise. The root mean square (RMS) error estimate relating to the standard deviation $\sigma$ of a Gaussian random variable obtained in the study \cite{carroll2018b} is of the form 
\begin{equation} 
\label{rms_estimate}
\sigma \approx \frac{1}{\sqrt{\tau}}  \, 300 \, \hbox{E}.  
\end{equation} 
Here $\tau$ refers to the duration of the measurement in seconds. The unit E is eotvos (1 E = 1e-9   $s^{-2}$). The RMS values of this study and the required measurement times calculated according to (\ref{rms_estimate}) can be found in Table \ref{table:noise_values}. The measurement durations for the higher noise levels shown in the Table \ref{table:noise_values} are unrealistically short from a measurement time perspective, as the shortest measurement time for the instrument is 30 seconds \cite{carroll2018b}. Therefore, they have been marked in parenthesis and are included for consistency.

The RMS noise level of the VEGA instrument is 55 E resulting in a measurement time of 27.4 seconds \cite{carroll2018b}. These finite measurement times follow from the repetitive measurement process required to measure the gradient of the gravitational field strength with a given accuracy.

\begin{table}[!ht]
    \centering
    \caption{The measurement noise levels investigated in this study. The level of the noise and the  duration required by the measurement have been estimated according to \cite{carroll2018b}. The measurement times in (c) and (d) are unrealistically short and have therefore been marked in parenthesis only for consistency.}
    \begin{tabular}{ccc}
    \hline
        Item & RMS (E)&  Duration (s) \\ 
        \hline
        (a) & 2  & 2.3E4 \\
        (b) & 8  & 1400  \\
        (c) & 80  & (14)  \\
        (d) & 800 & (0.14)  \\
    \end{tabular}
    \label{table:noise_values}
\end{table}

From the inversion viewpoint, the measurement noise constitutes one significant error source,  the other one being the modelling uncertainty related to, e.g., the position and orientation of the measurement instrument as well as the discretization of the computation domain. The methods for treating the positional error in this work are  discussed more in the section \ref{sec:modelling_error} which shows that the errors relating to the positional uncertainty can be decreased by a randomization and averaging procedure in the inversion stage.

\subsection{Forward model}\label{sec:forwardmodel}

The relationship between a given mass density and the measurement data can be presented via a linear forward model of the form
\begin{equation} 
\label{linear_system}
{\bf y} = {\bf L} {\bf x} + {\bf n},
    \end{equation}
in which ${\bf x} \in \mathbb{R}^N$ is the difference between the actual mass density, i.e., the unknown of the inverse problem, and an initial guess;  ${\bf y} \in \mathbb{R}^M$ represents the difference between the actual measurements and  numerically simulated data obtained for the initial  mass density; ${\bf L}$ is the system matrix representing the forward map;  and ${\bf n} \in \mathbb{R}^M$ is the noise vector.

We assume that the target has been decomposed into a finite number of disjoint elements $T_1, T_2, \ldots, T_m$ (here: tetrahedra), and that the mass density is a piecewise constant distribution with respect to a set of  disjoint subsets denoted by $R_{1}, R_{2}, \ldots, R_{N}$  consisting of $K_{j_1}, K_{j_2}, \ldots, K_{j_N}$ elements, respectively, with  $m = \sum_{j  = 1}^{N} K_j$. The subset $R_{j}$ is defined as the union of the elements with center of mass closest to the point $\vec{z}_j$ in the set $\vec{z}_1, \vec{z}_2, \ldots, \vec{z}_N$ of randomly generated points. 

The characteristic function of the element $T_j$, which is  equal to one within the set $j$ and zero elsewhere, is denoted by $\chi_j$. The resulting mass difference density, $\Delta \rho$, is of the form 
\begin{equation} 
\label{mass_density}
\Delta \rho = \sum_{j = 1}^{N} x_j \psi_j \quad \hbox{with} \quad \psi_j = \sum_{k_j = 1}^{K_j}  \chi_j. 
\end{equation} 

The $\psi_1, \psi_2, \ldots, \psi_N$ is the function basis of the mass difference density, which is assumed to have a piecewise constant distribution inside the target asteroid. It is also the function basis for the  inversion process.

Our data corresponds to the gradient of the scalar gravity field strength, i.e.,  the Euclidean norm of the three-component gravity field at a given point. The measurements are assumed to contain additive zero-mean Gaussian noise. Furthermore, the noise entries are assumed to be independent and identically distributed. The resulting forward model is of the form
\begin{eqnarray}
\label{measurement}
    y_i &= & G \int_\Omega  \Delta \rho({\vec z}) \nabla_{\vec {r}_i}  \frac{1}{\| {\vec z} - {\vec {r}_i } \|^2 } \, \hbox{d} {\vec z} + { n} \nonumber \\ & = & 2 G \int_\Omega  \Delta \rho({\vec z}) \frac{{\vec z} - {\vec {r}_i}}{\| {\vec z} - {\vec {r}_i} \|^4 } \, \hbox{d} {\vec z} + {n}_i, 
\end{eqnarray}
where $y_i$ with $i = 1, 2, \ldots, M$ represents the difference data at the point ${\vec r}_i$; $n_i$ is a noise vector; and $G$ is Newton's gravitational constant. 

Substituting (\ref{mass_density}) into (\ref{measurement}) and evaluating the gradient  results in the equation 
\begin{eqnarray}
    {y}_i & = & 2 G \sum_{j = 1}^{N}  x_j  \int_\Omega  \psi_j  \frac{{\vec  z} -   {\vec {r}}_i}{\| {\vec z} - {\vec {r}}_i \|^4 } \, \hbox{d} {\vec z} + {n}_i \nonumber \\ &=& 2 G \sum_{j = 1}^{N}  x_j  \sum_{k_j = 1}^{K_j}  \frac{{\vec  z}_{k_j} - {\vec {r}}_i}{\| {\vec z}_{k_j} - {\vec {r}}_i \|^4 } \, \int_{T_{k_j}} \hbox{d} {\vec z} + {n}_i, 
\end{eqnarray}
where $\vec{z}_{k_j}$ is the center of mass and $\int_{T_j} \hbox{d} {\bf z}$ the volume of $T_j$. The resulting matrix equation is of the form (\ref{linear_system}) with the matrix ${\bf L} \in \mathbb{R}^{M \times N}$ given by 
\begin{equation}
 L_{i,j} =   2 G  \sum_{k_j = 1}^{K_j} \frac{{\bf z}_{k_j} - {\bf {r}}_i}{\| {\bf z}_{k_j} - {\bf {r}}_i \|^4 } \, \int_{T_j} \hbox{d} {\bf z}. 
\end{equation}

\subsection{Inversion process} \label{sec:inversionprocess}

We approach the inverse problem via the hierarchical Bayesian model (HBM) in which the unknown parameter ${\mathbf x}$ obeys a {\em posterior} probability density determined by the product $p ( {\mathbf x}, {\bm \theta} \mid {\mathbf y})   \propto  {p({\mathbf x}, {\bm \theta}) \, p({\mathbf y} \mid {\mathbf x})}$ between the prior density $p({\mathbf x}, {\bm \theta})$, and the likelihood function $p({\mathbf y} \mid {\mathbf x})$. The prior is a joint density of the form $p ( {\mathbf x}, {\bm \theta}) \propto  p({\bm \theta}) \,  p ({\mathbf x} \mid {\bm \theta})$, 
where the conditional part $p ( {\mathbf x} \mid {\bm \theta})$ is a zero mean Gaussian density  with a diagonal covariance matrix predicted by the hyperprior $p ({\bm \theta})$. The hyperprior is long-tailed, meaning that ${\bf x}$ is likely to be a sparse vector with only few entries differing from zero. For this prior structure, HBM is advantageous for finding a well-localized reconstruction. As a hyperprior, one can use either the gamma  or  inverse gamma  density \cite{calvetti2009}, whose shape and scale are controlled by the parameters $\beta$ and $\theta_0$, respectively. The likelihood follows directly from the measurement noise density via ${\bf n} = {\bf y} - {\bf L} {\bf x}$ with independent entries. 

\subsubsection{IAS inversion}

The inverse estimate is found using the IAS {\em maximum a posteriori} (MAP) estimation method \cite{calvetti2009,calvetti2008,pursiainen2013}. The gamma density is applied as the hyperprior. The IAS algorithm finds the MAP estimate via alternating the conditional posteriors  $p ( {\bf x} \mid  {\bm \theta}, {\bf y}) $ and  $p ( {\bm \theta} \mid {\bf x}, {\bf y}) $ as the objective function. This is advantageous, since the maximizer for the first one can be found by solving a regularized least-squares optimization problem, and for the second one via an explicit analytical formula. When IAS inversion is applied with gamma density as the hyperprior, the outcome of the algorithm can be shown to correspond to the classical $\ell^2$-norm regularized solution of the inverse problem. 

\subsubsection{Randomized multiresolution scan}
\label{sec:modelling_error}
\begin{figure}[!ht]
\centering
    \includegraphics[width=0.225\textwidth]{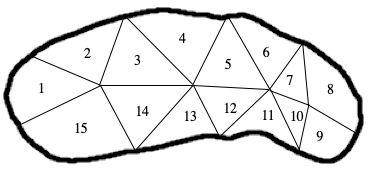}
    \includegraphics[width=0.225\textwidth]{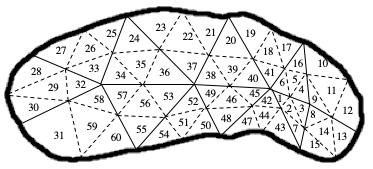}
    \caption{A schematic illustration of subdivision of the asteroid shape into coarse (left) and fine (right) resolution subdomains. In this case the sparsity factor $s$, the ratio between the number of subdomains for two consecutive resolution levels, is four (4). An example of a surjective mapping from the coarse to fine resolution, $\mathbb{R}^N \to \mathbb{R}^K_\ell$, is given by $\{2\} \to \{ 25,26,27,33\}$.}
    \label{fig:resolution_schematic}
\end{figure}

In order to minimize the effect of the selected function basis  $\psi_1, \psi_2, \ldots, \psi_N$ on the final reconstruction, we use a randomized  multiresolution scan algorithm which finds the final reconstruction $\overline{\overline{\bf x}}$  as an average ${\bf x}$-component of MAP estimates for ${\bm \zeta} = ({\bf x}, \theta)$ obtained for multiple different resolution levels (Figure \ref{fig:resolution_schematic}) and randomized decompositions of the parameter space. Each decomposition is  formed by selecting uniformly distributed number of points other than the selected mesh points with a nearest neighbor interpolation. Figure \ref{fig:resolution_schematic} shows schematically one possible randomized decomposition for two resolution levels. 

\begin{figure}[!ht]
\centering
    \includegraphics[width=5.5cm]{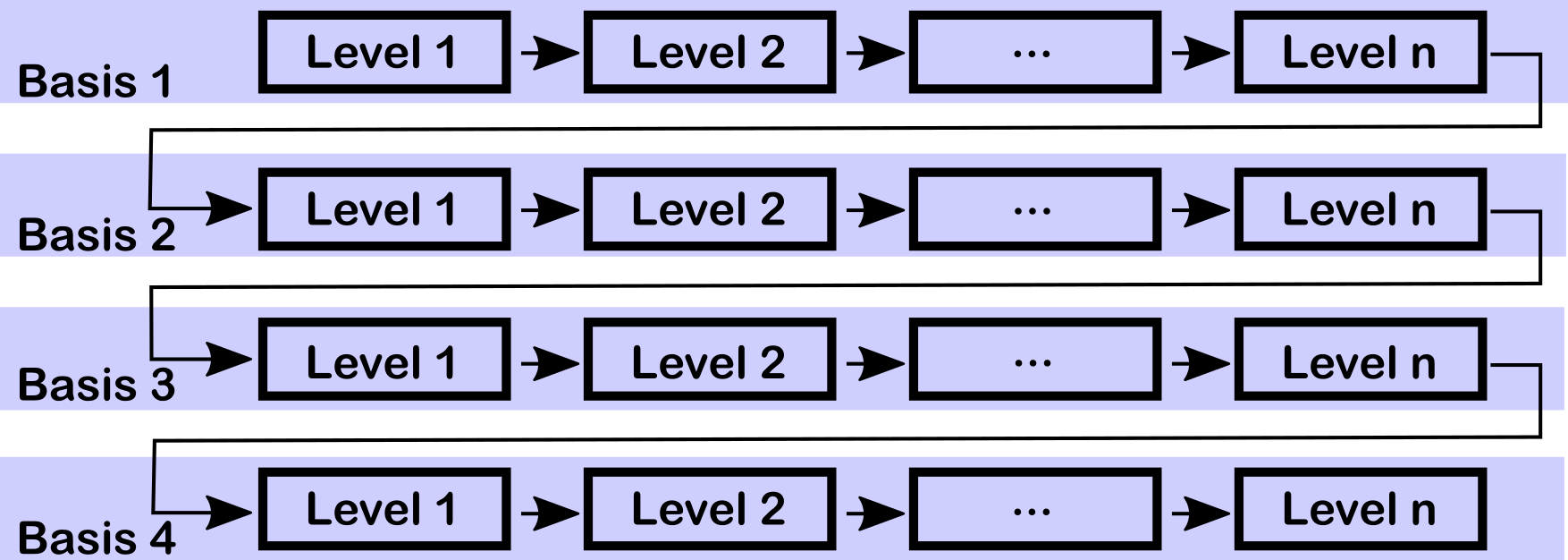}
    \caption{A schematic visualization of the data flow in the randomized multiresolution scan. The average of the MAP estimates obtained for the  $k$-th multiresolution decomposition is used as the  initial guess for the $k+1$-th one. }
    \label{fig:update_scheme}
\end{figure}

A decomposition $D_\ell$ refers to a surjective mapping $\mathbb{R}^N \to \mathbb{R}^K_\ell$ which is obtained by associating each basis function of the parameter space with the closest point in a set of $K_\ell$ random uniformly distributed points $\vec{p}_1, \vec{p}_2, \ldots, \vec{p}_{K_\ell}$. Each decomposition is organized into a sequence of subsets $S = \{B_1, B_2, \ldots, B_L \}$ in which the resolution, i.e., the number of randomized points grows along with $\ell$ as given by $K_\ell = K s^{(\ell - L) }$, where $s$ is a user-defined sparsity factor. Such a sequence is referred here to as a randomized multiresolution decomposition (Figure \ref{fig:resolution_schematic}). For optimizing the performance of the MAP estimation process, it is essential to begin with a coarse resolution and gradually proceed towards a finer one, since the distinguishability of the coarse density fluctuations which are realistic in asteroids is generally superior to that of other details such as small and well-localized density changes. The MAP estimate for $B_\ell+1$ is obtained by using the one for $B_\ell$ as the initial guess (Figure \ref{fig:update_scheme}). Analogously, the estimate obtained for a single subset sequence is used as the initial guess for the next one. The initial guess of the whole procedure is set to be ${\bf x}^{(0)} = (0, 0, \ldots, 0)$ and ${\bf \theta}^{(0)} = (\theta_0, \theta_0, \ldots, \theta_0)$. 

The algorithm proceeds as follows:
\begin{enumerate}
\item Choose the desired number of the resolution levels $L$ and the sparsity factor $s$, i.e., the ratio of source counts between the levels. 
\item For each resolution level $\ell = 1, 2, \ldots, L$, create a random uniformly distributed set of center points $\vec{p}_1, \vec{p}_2, \ldots, \vec{p}_{K_\ell}$. Find the sequence of subsets $B_1$, $B_2$, $\ldots$, $B_{K_\ell}$ by applying the nearest interpolation scheme with respect to the center points. 
\item Repeat the first two steps to generate a desired number $D$ of independent randomized multiresolution decompositions $S_1$, $S_2$, $\ldots$, $S_D$. 
\item  Start the reconstruction process with the decomposition $S_1$ and an initial guess ${\bm \zeta}^{(0)}$ corresponding to ${\bf x}^{(0)} = (0, 0, \ldots, 0)$ and ${\bf \theta}^{(0)} = (\theta_0, \theta_0, \ldots, \theta_0)$.
\item  For decomposition $S_k$,  find a reconstruction $x^{(\ell)}$ with an inversion technique chosen by the user, here the IAS method, and the initial guess ${\bm \zeta}^{(\ell-1)}$ for the resolution levels $\ell = 1, 2, \ldots, L$.  
\item For level $L$, obtain the final estimate for the decomposition (basis) $k$ as the normalized mean 
\begin{equation}  \overline{\bm \zeta}^{(k)} =   \sum_{\ell = 1}^L {\bm \zeta}^{(\ell)} \,  /  \,  \sum_{\ell = 1}^L  s^{(L - \ell)}. \end{equation} 
Here the denominator follows from balancing out the effect of the multiplied source count which follows from the interpolation of a coarse level estimate to a denser resolution level. 
\item If $k < D$  update $k \to k + 1$, i.e. move to the next decomposition, and repeat the 5-th and 6-th step with the initial guess $\overline{\bm \zeta}^{(k-1)}$ for the resolution level $\ell = 1$.
\item Obtain the final reconstruction $\overline{\overline{\bf x}}$ as the ${\bf x}$-component of the mean: 
\begin{equation}
\overline{\overline{\bm \zeta}}^{(k)} = \frac{1}{D} \sum_{k=1}^D \overline{\bm \zeta}^{(k)}.  
    \end{equation}
\end{enumerate} 

This procedure ensures that the $\ell^2$ solution converges to the solution which is independent of the discretization of the computation domain and makes the solution to converge into the $\ell^2$-norm regularized solution. Assuming that the possible density fluctuations due to the basis selection are random and identically distributed, the average should converge towards a reconstruction which is invariant with respect to the applied basis. Hence, the randomized multiresolution decomposition enables localization of density fluctuations of various sizes within the deep interior part of the domain. Using simply the fine mesh shows any density anomalies only on the surface of the reconstruction. The coarser mesh decomposition and randomized scanning algorithm average out errors due to discretization arising in the decomposition process.  

\subsection{Numerical implementation with Geoceles interface}

The present forward and inverse algorithms can be found implemented in the open Geoceles software package \cite{geoceles} for the Matlab (Mathworks, Inc.) platform. The software was developed over the course of this study and was applied in the numerical experiments. Geoceles creates a regular tetrahedral mesh conforming to the segmentation determined by the closed triangular surfaces. It applies graphics processing unit (GPU) acceleration which is essential in order to reduce computation time with reasonably high tetrahedral mesh resolution.

\begin{table*}[!ht]
    \centering
    \caption{Specifications of the features (i)--(viii) compared in the numerical experiments.}
    \begin{tabular}{llrrrr}
    \hline
         Item & Parameter &  \multicolumn{4}{c}{Values} \\
        \hline
        (i) & Noise & (a): 2 E  & (b): 8 E & (c): 80 E & (d): 800 E \\
        (ii) & Orbit radius & 500 m & 305 m \\
        (iii) & Angular coverage & 70$\mbox{}^\circ$ &  30$\mbox{}^\circ$  \\
        (iv) & IAS iterations $n_{\hbox{\scriptsize IAS}}$ & 1 & 5 & 10 \\
         & (Estimate type) & $\ell^2$ & $\ell^1$ \\ 
        (v) & Sparsity factor $s$ & 8 & 4 & 0 \\
(vi) & Number of decompositions & 100 & 10 & 1 \\
(vii) & Number of resolutions & 3 & 1  \\
 (viii) & Anomaly density & \multicolumn{2}{c}{Low (0 g/cm$^3$)} & \multicolumn{2}{c}{High (8.0 g/cm$^3$)} \\
 (ix) & Positional uncertainty & \multicolumn{2}{c}{5$^\circ$} & \multicolumn{2}{c}{ 10$^\circ$} \\
    \end{tabular}
    \label{table:specifications}
\end{table*}

\begin{table*}[!ht]
    \centering
    \caption{Numerical experiments ({\bf 1})--({\bf 4}).}
    \begin{tabular}{llll}
    \hline
         Experiment & Title  & Features investigated \\
        \hline
        ({\bf 1}) & Measurement configuration& (ii), (iii), (viii)  \\
        ({\bf 2}) & Noise level & (i), (ii) \\
        ({\bf 3}) & Multiresolution scanning & (ii), (vi)   \\
        ({\bf 4}) & Estimate type & (iv), (v), (vii) \\
    \end{tabular}
    \label{table:numerical_experiments}
\end{table*}

\subsection{Numerical experiments}

In the numerical experiments, we investigated the effects of the (i) noise, (ii) measurement distance, (iii) point distribution (angular coverage), (iv)  number of IAS iteration steps, (v) the resolution (sparsity factor), (vi)  number of the multiresolution decompositions and (vii) resolution levels, as well as (viii)  anomaly density. The specifications of the features (i)--(vii) can be found in Table \ref{table:specifications}. The numerical experiments were organized into four entities devoted to ({\bf 1}) orbit, ({\bf 2}) noise, ({\bf 3}) number of multiresolution decompositions,  as well as ({\bf 4}) resolution and estimate type. These focused on different features as described in Table \ref{table:numerical_experiments}. 

The likelihood standard deviation was selected to match with that of the relative standard deviation of the gravity gradient field magnitude. Gamma distribution was used as the hyperprior in the inversion process. The shape parameter $\beta$ was set to $1.5$ based on preliminary experimentation with the prior in these types of inversion tasks, and the scaling parameter $\theta_0$ was chosen based on the visual inspection of reconstruction quality (location and size of the detected anomaly in the reconstruction). The values were normalized with respect to the maximum entry. The workable values of $\theta_0$ at the lower noise levels were found to be $10^5$ and $10^8$ for the 305 and 500 meter radii orbits, respectively. These values produced an appropriate reconstruction quality in comparison to the exact model. The higher value was required for the high noise and low resolution cases. For the very low resolution, high noise, and higher orbit cases the scaling parameter had to be adjusted to $10^{10}$ in order to obtain a reasonable reconstruction showing the anomaly. The value of $\theta_0$ affects the shape of the prior and, consequently, also the MAP estimates. The higher values enhanced the reconstruction  and, therefore, they were used in the inversion process.

%% file: results.tex
\begin{figure*}[!ht] 
\centering
\begin{scriptsize}
\centering
305 m measurement distance  \\ \mbox{} \\
\begin{minipage}{0.243\textwidth}
\centering 
\includegraphics[width=\textwidth]{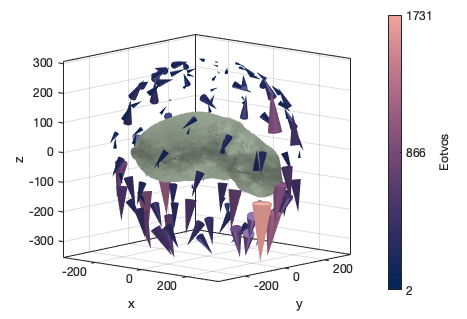} \\
Model (A) data \\ Orbit limiting angle 70$^\circ$ 
\end{minipage}
\begin{minipage}{0.243\textwidth}
\centering 
\includegraphics[width=\textwidth]{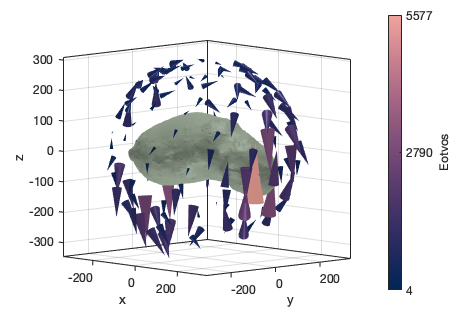}  \\
Model(B) data \\ Orbit limiting angle 70$^\circ$ 
\end{minipage} 
\begin{minipage}{0.243\textwidth}
\centering 
\includegraphics[width=\textwidth]{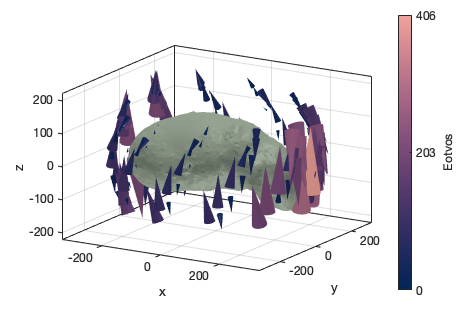} \\
Model (A) data \\ Orbit limiting angle 30$^\circ$ 
\end{minipage}
\begin{minipage}{0.243\textwidth}
\centering 
\includegraphics[width=\textwidth]{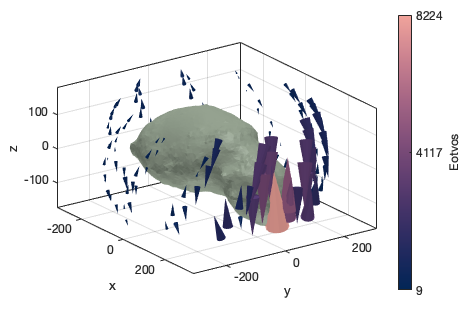} \\
Model (B) data \\  Orbit limiting angle 30$^\circ$ 
\end{minipage}
\\ \vskip0.5cm
\centering
500 m measurement distance  \\ \mbox{} \\
\begin{minipage}{0.243\textwidth}
\centering 
\includegraphics[width=\textwidth]{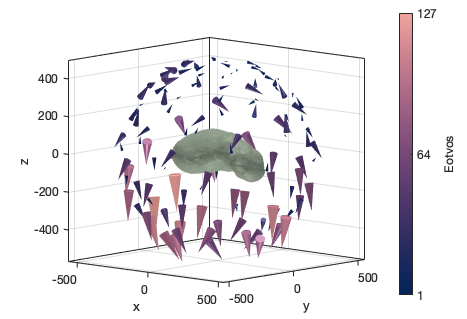}
Model (A) \\ Orbit limiting angle 70$^\circ$ 
\end{minipage}
\begin{minipage}{0.243\textwidth}
\centering 
\includegraphics[width=\textwidth]{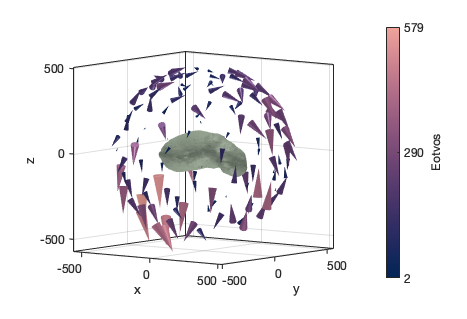} \\ Model (B) \\ Orbit limiting angle 70$^\circ$ \\
\end{minipage} 
\begin{minipage}{0.243\textwidth}
\centering 
\includegraphics[width=\textwidth]{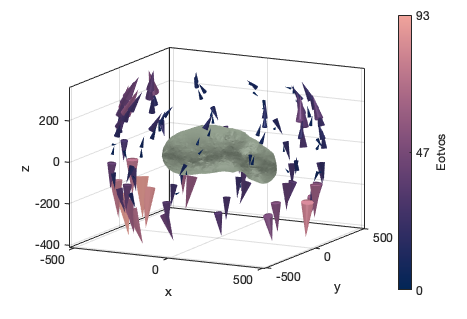} 
Model (A) \\ Orbit limiting angle 30$^\circ$ 
\end{minipage}
\begin{minipage}{0.243\textwidth}
\centering 
\includegraphics[width=\textwidth]{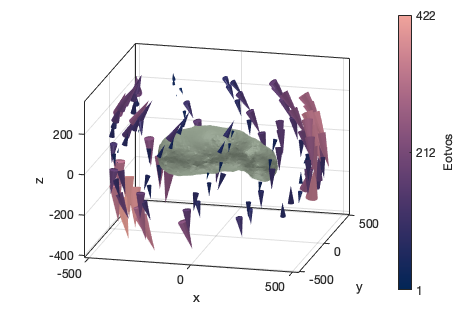} 
Model (B) \\ Orbit limiting angle 30$^\circ$ 
\end{minipage}

\end{scriptsize}
\caption{The magnitude and direction of the scalar gravity field gradient at the measurement points depicted for each measurement configuration and both measurement distances. }
\label{fig:measurement_data}
\end{figure*}

\begin{figure*}[!ht] 
\centering
\begin{scriptsize}
\centering
Projected data ${\bf L}{\bf x}^\ast$ for  500 m measurement distance  \\ \mbox{} \\
\vskip0.2cm
\begin{minipage}{0.243\textwidth}
\centering 
\includegraphics[width=\textwidth]{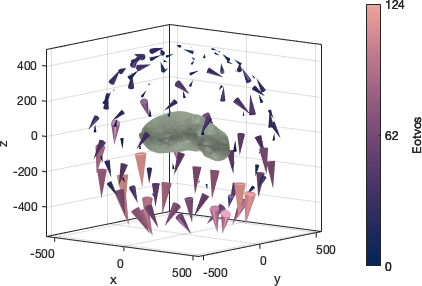}
Model (A) \\ Orbit limiting angle 70$^\circ$ 

\end{minipage}
\begin{minipage}{0.243\textwidth}
\centering 
\includegraphics[width=\textwidth]{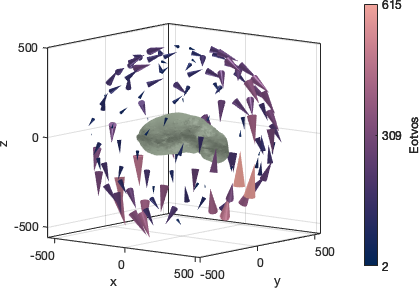} \\
Model (B) \\ Orbit limiting angle 70$^\circ$ 
\end{minipage} 
\begin{minipage}{0.243\textwidth}
\centering 
\includegraphics[width=\textwidth]{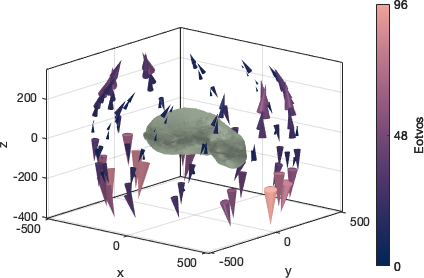} \\
Model (A) \\ Orbit limiting angle 30$^\circ$ 

\end{minipage}
\begin{minipage}{0.243\textwidth}
\centering 
\includegraphics[width=\textwidth]{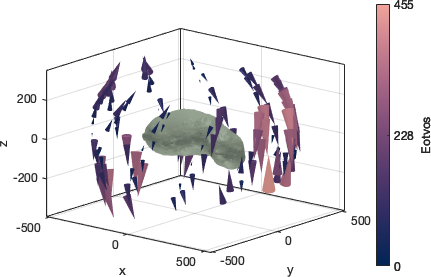} \\
Model (B) \\ Orbit limiting angle 30$^\circ$ 

\end{minipage}
\vskip0.5cm
Residual ${\bf y} - {\bf L}{\bf x}^\ast$ for  500 m measurement distance \\
\vskip0.2cm
\begin{minipage}{0.243\textwidth}
\centering 
\includegraphics[width=\textwidth]{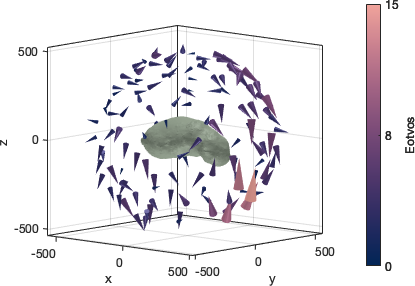} \\
Model (A) \\ Orbit limiting angle 70$^\circ$ 
\end{minipage}
\begin{minipage}{0.243\textwidth}
\centering 
\includegraphics[width=\textwidth]{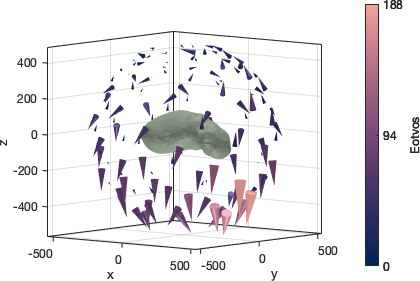} \\
Model (B) \\ Orbit limiting angle 70$^\circ$ 
\end{minipage} 
\begin{minipage}{0.243\textwidth}
\centering 
\includegraphics[width=\textwidth]{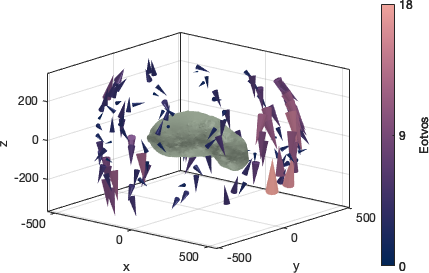} \\
Model (A) \\ Orbit limiting angle 30$^\circ$ 
\end{minipage}
\begin{minipage}{0.243\textwidth}
\centering 
\includegraphics[width=\textwidth]{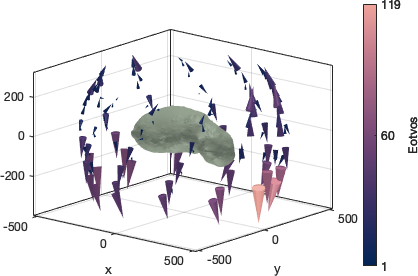} \\
Model (B) \\ Orbit limiting angle 30$^\circ$ 
\end{minipage}
\end{scriptsize}
\caption{The projected data ${\bf L}{\bf x}^\ast$ and the residual ${\bf y} - {\bf L}{\bf x}^\ast$ for the 500 m measurement distance. Here the reconstruction has been normalized so that the 2-norm of the actual and projected data coincide.  }
\label{fig:projected_and_residual}
\end{figure*}


\begin{figure*}[!ht] 
\centering
\begin{scriptsize}
\centering
305 m measurement distance  \\ \mbox{} \\
\begin{minipage}{0.243\textwidth}
\centering 
\includegraphics[width=\textwidth]{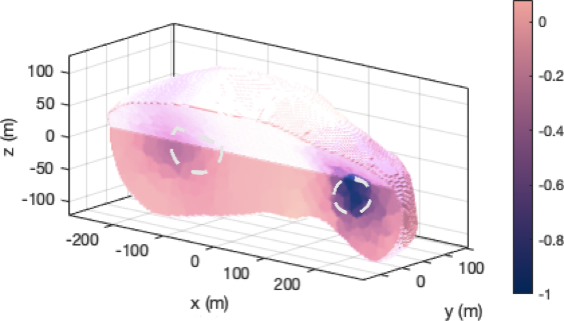} \\
Model (A) \\ Orbit limiting angle 70$^\circ$ 
\end{minipage}
\begin{minipage}{0.243\textwidth}
\centering 
\includegraphics[width=\textwidth]{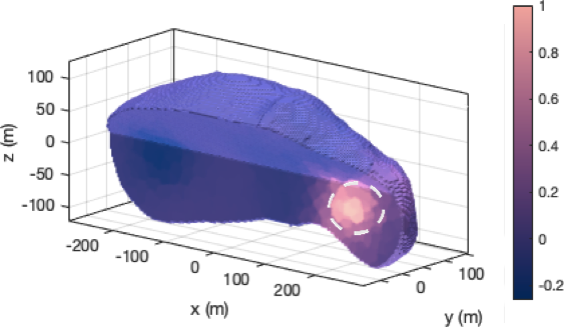} \\
Model(B) \\ Orbit limiting angle 70$^\circ$ 
\end{minipage} 
\begin{minipage}{0.243\textwidth}
\centering 
\includegraphics[width=\textwidth]{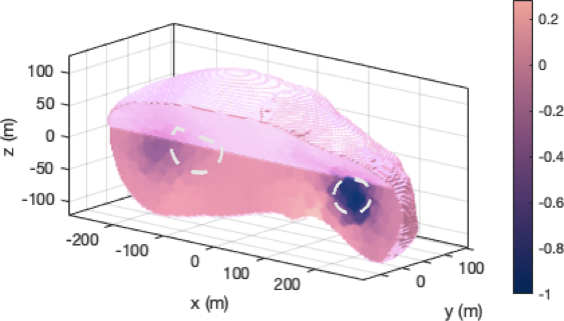} \\
Model (A)  \\ Orbit limiting angle 30$^\circ$ 
\end{minipage}
\begin{minipage}{0.243\textwidth}
\centering 
\includegraphics[width=\textwidth]{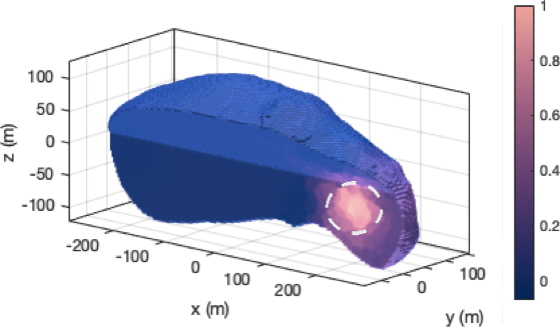} \\
Model (B) \\ Orbit limiting angle 30$^\circ$ 
\end{minipage}
\\ \vskip0.5cm
\centering
500 m measurement distance  \\ \mbox{} \\
\begin{minipage}{0.243\textwidth}
\centering 
\includegraphics[width=\textwidth]{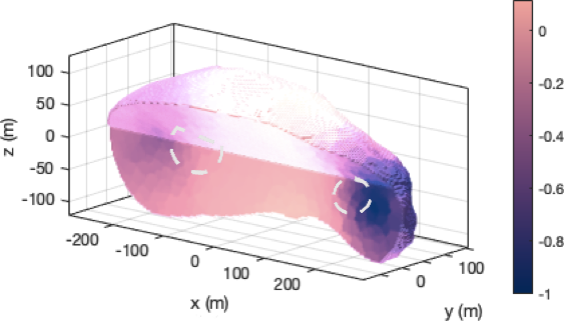}\\
Model (A) \\ Orbit limiting angle 70$^\circ$ 
\end{minipage}
\begin{minipage}{0.243\textwidth}
\centering 
\includegraphics[width=\textwidth]{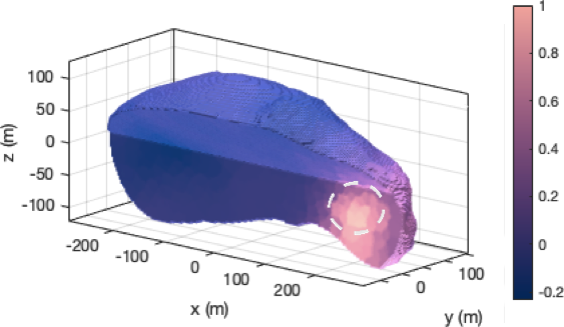}\\
Model (B) \\ Orbit limiting angle 70$^\circ$ 
\end{minipage} 
\begin{minipage}{0.243\textwidth}
\centering 
\includegraphics[width=\textwidth]{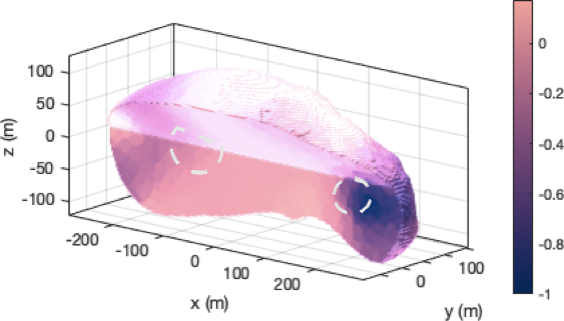} \\
Model (A) \\ Orbit limiting angle 30$^\circ$ 
\end{minipage}
\begin{minipage}{0.243\textwidth}
\centering 
\includegraphics[width=\textwidth]{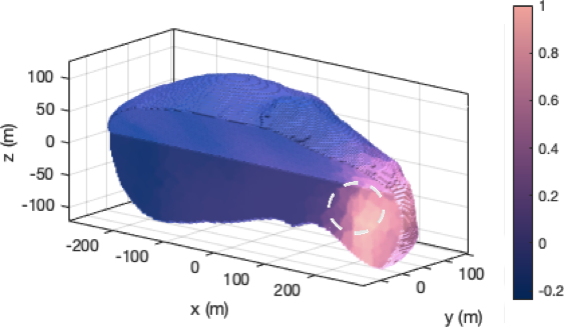} \\
Model (B) \\ Orbit limiting angle 30$^\circ$ 
\end{minipage}
\end{scriptsize}
\caption{3D cut-in views of the reconstructions at 305 and 500 m measurement distances and two different angular coverages with the corresponding measurement points shown in the Figure \ref{fig:measurementpoints}. The higher coverage of points around the asteroid (limiting angle 70$^\circ$) results in better localization of the perturbation at the higher orbit radius. With the smaller radius the effect on the reconstructions is not significant. The exact locations of the density anomalies are indicated with dashed white circles.}
\label{fig:perturbations}
\end{figure*}


\begin{figure*}[!ht] 
\centering
\begin{scriptsize}

\centering
500 m measurement distance  \\ \mbox{} \\
\begin{minipage}{0.243\textwidth}
\centering 
\includegraphics[width=\textwidth]{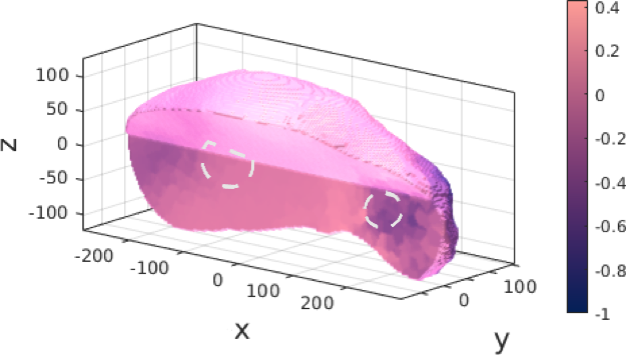}\\
Model (A) \\ Rotation 5 degrees 
\end{minipage}
\begin{minipage}{0.243\textwidth}
\centering 
\includegraphics[width=\textwidth]{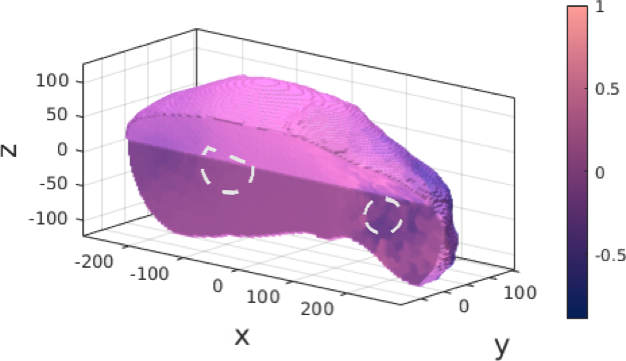}\\
Model (A) \\ Rotation 10 degrees 
\end{minipage} 
\begin{minipage}{0.243\textwidth}
\centering 
\includegraphics[width=\textwidth]{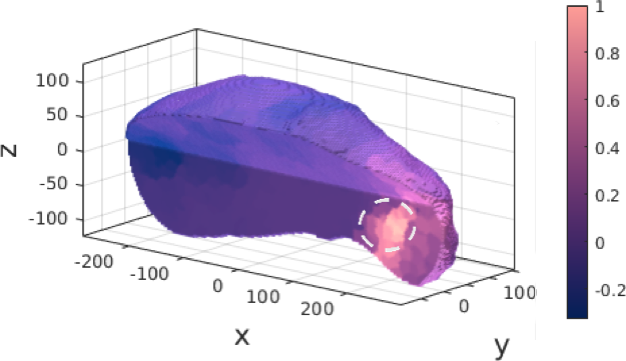}\\
Model (B) \\ Rotation 5 degrees 
\end{minipage}
\begin{minipage}{0.243\textwidth}
\centering 
\includegraphics[width=\textwidth]{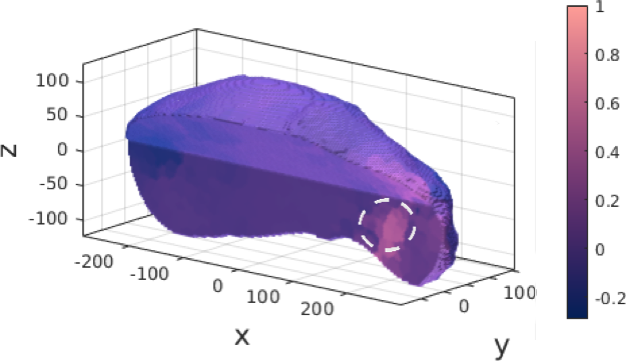}\\
Model (B) \\ Rotation 10 degrees 
\end{minipage}
\end{scriptsize}
\caption{The effect of positional uncertainty on the reconstructions. Such uncertainty results from measurements being carried out on a measurement arc instead of points. The exact locations of the density anomalies are indicated with dashed white circles.}
\label{fig:arc_rotation}
\end{figure*}


\begin{figure*}[!ht]
\begin{tiny}
\centering 
Model (A), 500 m measurement distance  \\ \mbox{} \\
\begin{minipage}{0.225\textwidth}
\centering
\includegraphics[width=0.80\textwidth]{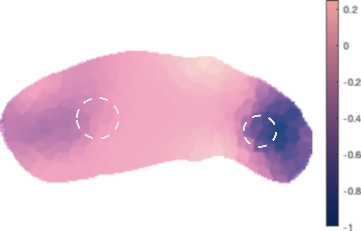} \\
$\sigma = 2$ E, $s = 8$, \\ $D = 100$,  $n_{\hbox{\tiny IAS}} = 1$
\end{minipage}
\begin{minipage}{0.225\textwidth} 
\centering
\includegraphics[width=0.80\textwidth]{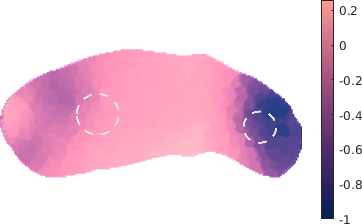} \\
$\sigma = 8$ E, $s = 8$, \\ $D = 100$,  $n_{\hbox{\tiny IAS}} = 1$
\end{minipage}
\begin{minipage}{0.225\textwidth} 
\centering
\includegraphics[width=0.80\textwidth]{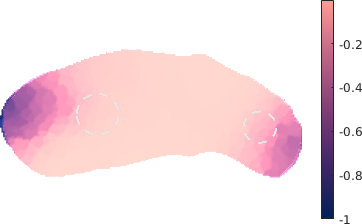} \\
$\sigma = 80$ E, $s = 8$, \\ $D = 100$,  $n_{\hbox{\tiny IAS}} = 1$
\end{minipage}
\begin{minipage}{0.225\textwidth} 
\centering
\includegraphics[width=0.80\textwidth]{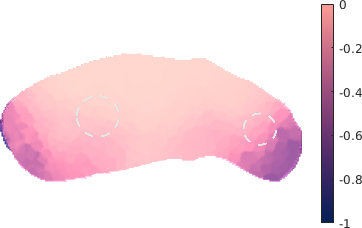} \\
$\sigma = 800$ E, $s = 800$, \\ $D = 100$,  $n_{\hbox{\tiny IAS}} = 1$
\end{minipage} \\
\begin{minipage}{0.225\textwidth} 
\centering
\includegraphics[width=0.80\textwidth]{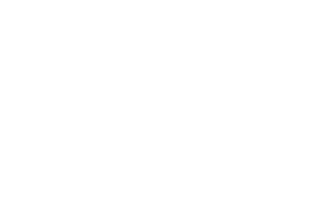}
\end{minipage}
\begin{minipage}{0.225\textwidth} 
\centering
\includegraphics[width=0.80\textwidth]{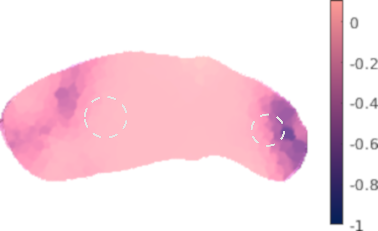}
\\
$\sigma = 2$ E, $s = 4$, \\ $D = 100$,  $n_{\hbox{\tiny IAS}} = 1$
\end{minipage}
\begin{minipage}{0.225\textwidth} 
\centering
\includegraphics[width=0.80\textwidth]{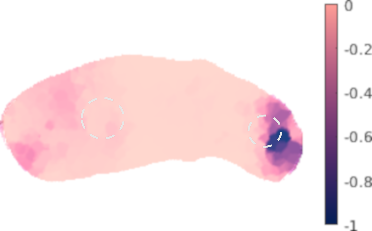}
\\
$\sigma = 2$ E, $s = 4$, \\ $D = 10$,  $n_{\hbox{\tiny IAS}} = 1$
\end{minipage}
\begin{minipage}{0.225\textwidth} 
\centering
\includegraphics[width=0.80\textwidth]{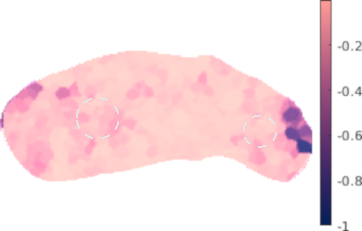}
\\
$\sigma = 2$ E, $s = 4$, \\ $D = 1$,  $n_{\hbox{\tiny IAS}} = 1$
\end{minipage} \\
\begin{minipage}{0.225\textwidth} 
\centering
\includegraphics[width=0.80\textwidth]{empty_305m.png}
\end{minipage}
\begin{minipage}{0.225\textwidth} 
\centering
\includegraphics[width=0.80\textwidth]{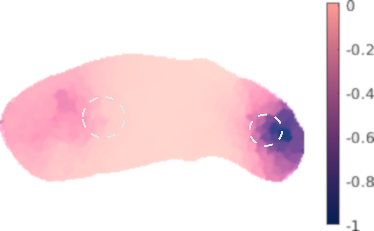}
\\
$\sigma = 2$ E, $s = 8$, \\ $D = 100$,  $n_{\hbox{\tiny IAS}} = 5$
\end{minipage}
\begin{minipage}{0.225\textwidth} 
\centering
\includegraphics[width=0.80\textwidth]{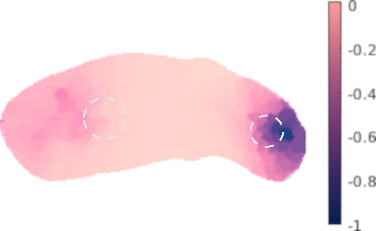}
\\
$\sigma = 2$ E, $s = 8$, \\ $D = 100$,  $n_{\hbox{\tiny IAS}} = 10$
\end{minipage}
\begin{minipage}{0.225\textwidth} 
\centering
\includegraphics[width=0.80\textwidth]{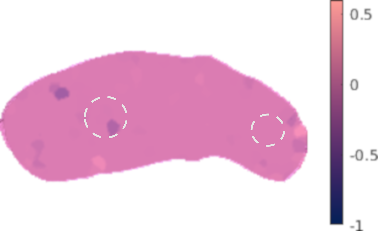}
\\
$\sigma = 2$ E, $s = 0$, \\ $D = 1$,  $n_{\hbox{\tiny IAS}} = 1$
\end{minipage}

 \vskip0.5cm
\centering 
Model (B), 500 m measurement distance  \\ \mbox{} \\
\begin{minipage}{0.245\textwidth}
\centering
\includegraphics[width=0.80\textwidth]{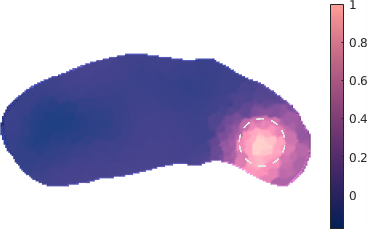} \\
$\sigma = 2$ E, $s = 8$, \\ $D = 100$,  $n_{\hbox{\tiny IAS}} = 1$
\end{minipage}
\begin{minipage}{0.245\textwidth} 
\centering
\includegraphics[width=0.80\textwidth]{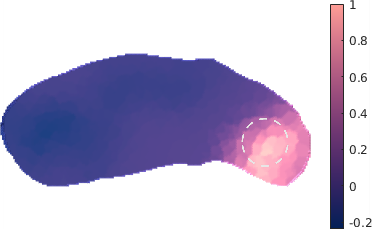} \\
$\sigma = 8$ E, $s = 8$, \\ $D = 100$,  $n_{\hbox{\tiny IAS}} = 1$
\end{minipage}
\begin{minipage}{0.245\textwidth} 
\centering
\includegraphics[width=0.80\textwidth]{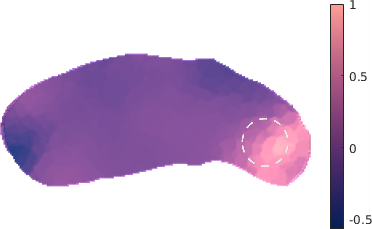} \\
$\sigma = 80$ E, $s = 8$, \\ $D = 100$,  $n_{\hbox{\tiny IAS}} = 1$
\end{minipage}
\begin{minipage}{0.245\textwidth} 
\centering
\includegraphics[width=0.80\textwidth]{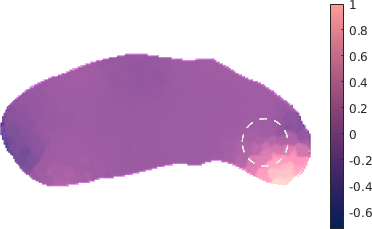} \\
$\sigma = 800$ E, $s = 8$, \\ $D = 100$,  $n_{\hbox{\tiny IAS}} = 1$
\end{minipage} \\
\begin{minipage}{0.245\textwidth} 
\centering
\includegraphics[width=0.80\textwidth]{empty_305m.png}
\end{minipage}
\begin{minipage}{0.245\textwidth} 
\centering
\includegraphics[width=0.80\textwidth]{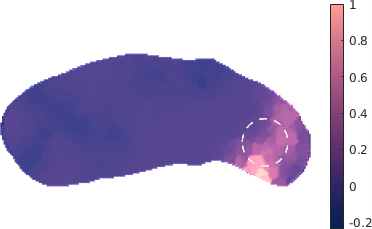}
\\
$\sigma = 2$ E, $s = 4$, \\ $D = 100$,  $n_{\hbox{\tiny IAS}} = 1$
\end{minipage}
\begin{minipage}{0.245\textwidth} 
\centering
\includegraphics[width=0.80\textwidth]{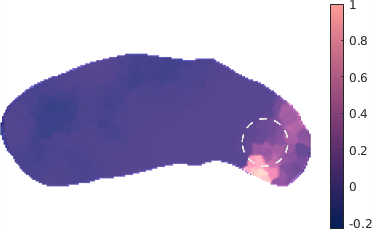}
\\
$\sigma = 2$ E, $s = 4$, \\ $D = 10$,  $n_{\hbox{\tiny IAS}} = 1$
\end{minipage}
\begin{minipage}{0.245\textwidth} 
\centering
\includegraphics[width=0.80\textwidth]{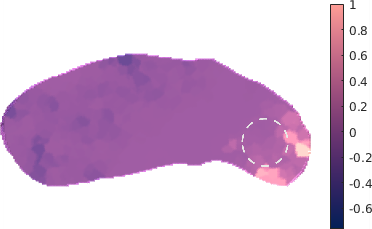}
\\
$\sigma = 2$ E, $s = 4$, \\ $D = 1$,  $n_{\hbox{\tiny IAS}} = 1$
\end{minipage} \\
\begin{minipage}{0.245\textwidth} 
\centering
\includegraphics[width=0.80\textwidth]{empty_305m.png}
\end{minipage}
\begin{minipage}{0.245\textwidth} 
\centering
\includegraphics[width=0.80\textwidth]{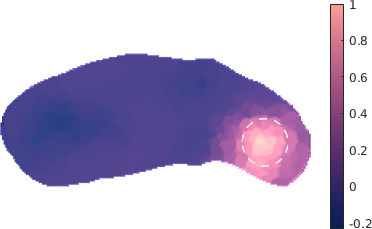}
\\
$\sigma = 2$ E, $s = 8$, \\ $D = 100$,  $n_{\hbox{\tiny IAS}} = 5$
\end{minipage}
\begin{minipage}{0.245\textwidth} 
\centering
\includegraphics[width=0.80\textwidth]{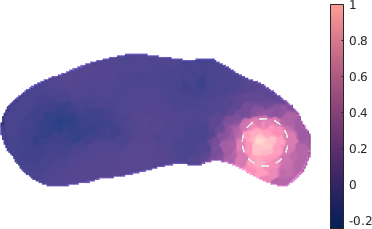}
\\
$\sigma = 2$ E, $s = 8$, \\ $D = 100$,  $n_{\hbox{\tiny IAS}} = 10$
\end{minipage}
\begin{minipage}{0.245\textwidth} 
\centering
\includegraphics[width=0.80\textwidth]{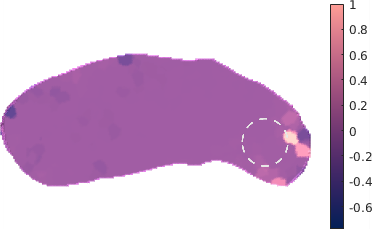}
\\
$\sigma = 2$ E, $s = 0$, \\ $D = 100$,  $n_{\hbox{\tiny IAS}} = 1$
\end{minipage}
\caption{The xz-plane crosscut views of the effects of noise level, sparsity factor, and IAS MAP iteration rounds $n_{\hbox{\tiny IAS}}$ on the Model (A) and (B) reconstructions at 500 m measurement distance. The white dashed lines indicate the locations of the density anomalies. Parameters $\sigma$, $s$ and $D$ denote the standard deviation of the noise, the sparsity factor and the number of the multiresolution decompositions, respectively. The number of resolution levels is $L=3$ in each case, and the number of IAS MAP iterations $n_{\hbox{\tiny IAS}}$ performed for each level is identical. Sparsity 0 refers to a case in which the multiresolution decomposition was absent in the reconstruction process.}
\label{fig:noiseresolutioniterationsmodelB}
\end{tiny}
\end{figure*}

\subsection{Measurement data}

The magnitude and direction of the simulated scalar gravity field gradient is shown in the Figure \ref{fig:measurement_data} for both measurement distances. 
The data shows clearly, how the strength of the field varies between the two models and the orbit distances, and that the difference is generally tangential. The high-density anomaly in the Model (B)  causes more than five times higher field strengths in comparison to the Model (A) which contained two voids inside the asteroid. The measurement distance has an effect on, not just the magnitude of the field strength,  but also on the localization of the anomalies. While there is clear localization of the signal in the lower orbit, similar effect is not seen as clearly in the more distant one, suggesting that a field originating from a distant source is averaged over the distance. The effect of the high-density anomaly in the Model (B) is clear in the 305 meter orbit case. The field strength for Model (A) is  generally more evenly distributed than for (B). However, that is to be expected as the mass distribution of (A) is more symmetric in comparison to (B). 

Figure \ref{fig:projected_and_residual} shows the projected data, which follows from multiplying the reconstruction with the system matrix, and also the residual, i.e., the difference between the actual and projected data, for the measurement distance of 500 m. The residual is obviously more random than the original data, which is in agreement with the current forward model in which the residual coincides with the noise term. The location of the largest amplitude is similar in each case, suggesting that the residual includes a model-driven component, which is not predicted by the present white noise error model. 

\subsection{Quality of reconstructions}

The quality of reconstructions was inspected visually by examining how well the different parts of the exact models (A) and (B) (Figure \ref{fig:exactmodel}) were visible in the reconstructions. The size, shape and location of any density anomalies indicated by the reconstructions was compared with the exact model. The good quality reconstructions showed clear, spherical anomalies of the size of the exact model, in the correct locations. The surface layer was not clearly distinguishable in any of the reconstructions.  

\subsection{Measurement configuration}

In the experiment ({\bf 1}), the visually assessed quality of the reconstruction was affected more by the distance between the target and the measurement points than the point coverage of the measurements, as shown by the Figure \ref{fig:perturbations} depicting the orbits with 305 and 500 meter radii. Distinguishing both low and high density anomalies, i.e., voids and boulders in the models (A) and (B), respectively, was found to be feasible at both distances. The closer, 305 m, orbit was observed to provide a superior depth localization capability, as the  reconstructions obtained at 500 m distance were somewhat biased towards the surface of the target domain, which is shown by the greater spread of the anomaly detected.

\subsection{Noise level}

The signal-to-noise ratio (SNR) for  Model (A) varied  from poor 12 dB to appropriate 24 dB, and  for Model (B) it was generally appropriate, ranging from 20 to 29 dB. The SNR for each case can be found in the Table \ref{table:snr}

\begin{table}[!ht]
    \centering
    \caption{Signal-to-noise ratios (SNR) for the investigated models. The SNR for the Model (A) varies from  poor to acceptable, whereas for the Model (B) it is generally appropriate.}
    \begin{tabular}{lccc}
    \hline
        $\theta$ & r (m) & Model (A) & Model (B) \\
        \hline
        70$^\circ$ & 305 & 24 dB & 28 dB \\
        & 500 & 14 dB & 21 dB \\
        30$^\circ$ & 305 & 21 dB & 29 dB \\
        & 500 & 12 dB & 20 dB \\
    \end{tabular}
    \label{table:snr}
\end{table}

The effect of the noise level was investigated for the 500 m orbit in the experiment ({\bf 2}) the results of which can be found in  Figure \ref{fig:noiseresolutioniterationsmodelB}. Anomaly detection  was found to be feasible with the lowest noise level (a) for Model (A) and (B). Above that, the voids found for Model (A) were barely distinguished as such. Detection of the high density anomaly in Model (B) was less affected by the noise.  However, the noise level (d) was too large for finding an appropriate reconstruction also in the case of (B).

\subsection{Measurement arc}

Realistic gravity gradient field measurements result in measurement arcs instead of points, causing positional uncertainty in the inversion stage. The effect of such uncertainty on the reconstruction is presented in the Figure \ref{fig:arc_rotation} for the 500 meters measurement radius. The reconstruction quality for Model (A) was found to  suffer from the positional uncertainty more than that found for  Model (B). 

\subsection{Multiresolution scanning}

The effects related to the number and sparsity factor of the multiresolution decompositions were explored in the experiment ({\bf 3}) and are illustrated in the Figure \ref{fig:noiseresolutioniterationsmodelB}. The smoothest results were obtained using 100 decompositions and the sparsity factor $s=8$. With a lower value $s=4$, i.e., with a smaller difference between the coarse and fine resolution levels, the reconstructions became biased towards the surface, losing depth-resolution, and regularity regarding the shape of the detected anomalies. The irregularity and bias were found to be further exaggerated, when a lower number of decompositions (10 and 1) were used. The reconstructions obtained with a single decomposition also included some visible high-frequency artifact patterns which, otherwise, were essentially absent. 

\subsection{Estimate type}

A comparison between $\ell^2$ and $\ell^1$ estimate types ({\bf 4}) can also be found in the Figure \ref{fig:noiseresolutioniterationsmodelB}. The  $\ell^2$ estimate, obtained with $n_{\hbox{\tiny IAS}} = 1$, provided, overall, the most regular (smooth) outcome. Increasing the number of IAS iterations, i.e., finding an $\ell^1$ type reconstruction, was observed to result in a sharper localization of the anomalies, sharpening also the other structures. The difference between the iteration counts of $n_{\hbox{\tiny IAS}} = 5$ and $n_{\hbox{\tiny IAS}} = 10$ was found to be a minor one. The essence of the multiresolution approach for the reconstruction quality is reflected by the single-resolution $\ell^2$-reconstruction ($n_{\hbox{\tiny IAS}} = 1$ with a single resolution level) which suffers from an extreme surface bias and artifacts.

%% file: discussion.tex
Direct measurements of asteroid interiors have yet not been carried out \cite{herique2017} and, therefore, our understanding of the internal structures are based on bulk properties, asteroid spin rates, as well as on impact and other simulation studies suggesting candidate examples of structures fitting to the parameters. For example, the asteroid Itokawa is known to have a 40 \% bulk porosity  \cite{Saito2006, sanchez2014} and a potential  aggregate structure \cite{barnouinjha}. The distribution of mass inside an asteroid is yet still mainly unknown. Our study suggests that the gradient of the gravity  field  can provide meaningful data of the internal distribution of mass, and that the data can be used for tomographic inversion. 

The two synthetic models developed for this study were selected based on the confirmed bulk parameters for the asteroid Itokawa, and on possible scenarios how these properties can be achieved, to provide baseline information on the kinds of deep interior structures that can be observed from the orbit. The results obtained in \cite{lowry2014} suggest that Itokawa is composed of the merger of two separate bodies with bulk densities of 1.750$\pm$0.110 g/cm$^3$ and 2.850$\pm$0.500 g/cm$^3$. They used the  shape model of  \cite{gaskell2008} which was based on rotational light curve analysis and applied detailed thermophysical analysis to the shape determined by the Hayabusa spacecraft \cite{Saito2006}. 

The double-void synthetic Model (A) corresponds to a  low-density asteroid with significant void space in the interior. The high-density anomaly contained by the Model (B) corresponds to one possible formation scenario of  \cite{lowry2014}, a catastrophic collision on a differentiated large object which contained a high-density metallic fragment (e.g. iron) in this formation process. This fragment could then then have been subsumed by the silicate material in the "head" region. Although it is unlikely that two unrelated objects would have a sufficiently low-velocity encounter to ensure the survival of both lobes and still have uniform surface composition and structure on the two lobes, and assumption of the existence of such iron-containing lobe has not been verified by direct measurements, the suggested model was included in this study to examine the reconstruction of such a high-density anomaly by gravity gradient field inversion.

The insensitivity of the depth localization for the measurement distance observed in this study is a well-known feature in ground based gravity inversion applications \cite{erkan2011, madej2017}. An obvious reason for this finding is the inherent ambiguity of depth information in gravity field data. Consequently, a nearby low-intensity anomaly can result in almost identical measurements with an outcome following from a higher-intensity and more distant obstacle. As a result, fluctuations in the depth (radial) direction are difficult to be reconstructed. That is, when a vector corresponding to such a fluctuation is multiplied with the governing matrix ${\bf L}$ of the forward model, the resulting vector is likely to belong into the numerical null-space of ${\bf L}$ \cite{liu1995, pursiainen2008}, meaning that it can have a norm very close to zero and making it weakly distinguishable based on the measurements. In the inversion stage, the weak depth-localization capability of the gravity data causes a strong bias of the reconstructions towards the surface of the target body which is why surface projections are used in presenting the results of gravity measurements, for example, the gravity acceleration on the surface of Mars in \cite{krzysztof2018}. The surface bias of a reconstruction is also a general phenomenon in inverse problems involving weak depth data, such as the biomedical imaging applications based on quasi-static electric field measurements \cite{kaipio2004,calvetti2009}. 

The essence of the proposed randomized multiresolution scan inversion approach is that the coarse density fluctuations are likely to be the clearest distinguishable components of the candidate solution set. Therefore, separating the coarse and fine details during the inversion process can provide an enhanced robustness of the final reconstruction compared to the approaches using all components at once. Here this is, especially, the case with correcting the bias in the depth direction. However, inverting measurements with a coarse discretization instead of a fine one might introduce other biases or artifacts relating to spatial location which in the present scanning approach are tackled by averaging the estimates produced by the IAS iteration over the set of randomized decompositions. The eventual inversion process can detect density details of various sizes as it can operate over a wide range of resolutions, and does not necessitate fixing the resolution {\em a priori} based on the deemed anomaly size. 

Obtaining data from a close enough distance will obviously require a special mission design. A close approach for gravity measurement has been suggested, e.g., for Juventas CubeSat in \cite{juventas}. Manouvering in the close proximity of the target will be difficult due to the inhomogeneities of the gravity field and, therefore, performing effective gravity field or gravity gradient measurements in the close proximity to a target SSSB will be challenging. Investigating realistic stable orbits around small bodies for gravimetric measurements will, therefore, be an important future research topic. A measurement duration of 5000 s resulting in a noise RMS of 4 E has been suggested to be feasible in \cite{carroll2018b}. Also incorporating the measurement arc instead of the point-approximation of the measurement used in this study and finding ways to incorporate model-driven components in the error model will be studied in the future. Obtaining a sufficiently low noise might be also possible via a radio-scientific Doppler shift measurement. Such approach was utilized, for example, in the high-precision measurements of the recent lunar science mission GRAIL (Gravity Recovery and Interior Laboratory) with two spacecraft configuration \cite{konopliv2013jpl,andrews2013ancient, zuber2013}. In addition to the signal specifications, the accuracy of a Doppler measurement will be limited also by external factors such as the solar wind pressure which presents a challenge with SSSBs of the size such as Itokawa and other asteroids with a few hundred meter diameters. 

Another important direction for the future work will be to investigate parallel radar and gravity inversion. Akin to the ground-based geoimaging applications \cite{erkan2011, mochales2008, hausmann2007}, a radar-based reconstruction of the interior structure may be expected to provide a superior depth-resolution compared to gravimetry, since the electromagnetic wave of the radar signal carries time-domain information which enables depth-inference of the scattering obstacle. Hence, a radar observation will be also less dependent on the observation distance in comparison to gravity measurements. Our recent findings  suggest that radar inversion emphasizes high-contrast details with the cost of smooth variations such as large areas of denser or more porous regolith which might be detected by a gravimeter \cite{sorsa2019}. Therefore, it will also be vital to analyze the interconnection and correlation between the electric permittivity and mass density distributions observed by the radar and gravimeter, respectively \cite{trabelsi2001}. Validating the gravity gradiometry inversion approach proposed in this paper with measurements in a terrestial application such as suggested by \cite{kirkendall2007} is also a potential future work topic.

%% file: conclusion.tex
This article concentrated on the mathematical methodology, resolution, noise and orbit radius selection with  tomographic gravity field investigation performed for an asteroid as the potential application. The results obtained with simulated data suggest that the randomized multiresolution scanning technique combined with the iterative alternating sequential (IAS) inversion algorithm provides an advantageous way to enable depth localization of density anomalies, a feature not inherent in other inversion methods. 

Void localization in an asteroid would require the measurement noise to be below 8 E at 500 meters orbit. The reconstruction of a high-density anomaly is less sensitive to noise, and it can be achieved with noise levels up to 80 E. This is also reflected in the signal-to-noise ratios obtained for the models. The reconstruction method was shown to be robust with respect to the positional uncertainty of the measurement for up to 5 \% error, confirming long enough measurement times to be able to reduce the measurement noise in real measurement scenarios. 